 \documentclass[preprint, twocolumn]{aastex6}

\newcommand{\lalpha}{LMg}   
\newcommand{\halpha}{HMg}   

\shorttitle{Chemical and Star Formation History for Halo Populations}
\shortauthors{Fern\'andez-Alvar et al.}

\begin{document}


\title{Disentangling the Galactic Halo with APOGEE: II. Chemical and Star Formation Histories for the Two Distinct Populations}


\author{Emma Fern\'andez-Alvar\altaffilmark{1}}

\author{Leticia Carigi\altaffilmark{1}}

\author{William J. Schuster\altaffilmark{2}}

\author{Christian R. Hayes\altaffilmark{3}}

\author{Nancy \'Avila-Vergara\altaffilmark{1,4}}

\author{Steve R. Majewski\altaffilmark{3}}

\author{Carlos Allende Prieto\altaffilmark{5,6}}

\author{Timothy C. Beers\altaffilmark{7}}

\author{Sebasti\'an F. S\'anchez\altaffilmark{1}}

\author{Olga Zamora\altaffilmark{5,6}}

\author{Domingo An\'ibal Garc\'ia-Hern\'andez\altaffilmark{5,6}}

\author{Baitian Tang\altaffilmark{8}}

\author{Jos\'e G. Fern\'andez-Trincado\altaffilmark{8,9}}

\author{Patricia Tissera\altaffilmark{10}}

\author{Douglas Geisler\altaffilmark{8}}

\and

\author{Sandro Villanova\altaffilmark{8}}             



\altaffiltext{1}{Instituto de Astronom\'{\i}a, 
              Universidad Nacional Aut\'onoma de M\'exico, AP 70-264, 04510, Ciudad de M\'exico,  M\'exico}
\altaffiltext{2}{Instituto de Astronom\'{\i}a, 
              Universidad Nacional Aut\'onoma de M\'exico, AP 106, 22800 Ensenada, B. C., M\'exico}
\altaffiltext{3}{Department of Astronomy, University of Virginia, Charlottesville, VA 22904-4325, USA}
\altaffiltext{4}{Departamento de F\'isica y Matem\'aticas, Universidad Iberoamericana, Prolongaci\'on Paseo de la Reforma 880, Lomas de Santa Fe, CP 01210 M\'exico DF, M\'exico}
\altaffiltext{5}{Instituto de Astrof\'{\i}sica de Canarias,
              V\'{\i}a L\'actea, 38205 La Laguna, Tenerife, Spain}
\altaffiltext{6}{Universidad de La Laguna, Departamento de Astrof\'{\i}sica, 
             38206 La Laguna, Tenerife, Spain}
\altaffiltext{7}{Department of Physics and JINA Center for the Evolution of the Elements,
University of Notre Dame, Notre Dame, IN 46556  USA}

\altaffiltext{8}{Departamento de Astronom\'ia, Casilla 160-CUniversidad de Concepci\'on, Concepci\'on, Chile}
\altaffiltext{9}{Institut Utinam, CNRS UMR6213, Univ. Bourgogne Franche-Comt\'e, OSU THETA , Observatoire de Besan\c{c}on, BP 1615, 25010 Besan\c{c}on Cedex, France }

\altaffiltext{10}{Departamento de Ciencias F\'isicas, Universidad Andres Bello, Av. Republica 220, Santiago, Chile}



\begin{abstract}

The formation processes that led to the current Galactic stellar halo are still under debate. Previous studies have provided evidence for different stellar populations in terms of elemental abundances and kinematics, pointing to different chemical and star-formation histories. In the present work we explore, over a broader range in metallicity ($-2.2 < \rm[Fe/H] < +0.5$), the two stellar populations detected in the first paper of this series from metal-poor stars in DR13 of the Apache Point Observatory Galactic Evolution Experiment (APOGEE). We aim to infer signatures of the initial mass function (IMF) and the star-formation history (SFH) from the two $\alpha$-to-iron versus iron abundance chemical trends for the most APOGEE-reliable $\alpha$-elements (O, Mg, Si and Ca). Using simple chemical-evolution models, we infer the upper mass limit ($M_{up}$) for the IMF and the star-formation rate (SFR), and its duration for each population. Compared with the low-$\alpha$ population, we obtain a more intense and longer-lived SFH, and a top-heavier IMF for the high-$\alpha$ population.

\end{abstract}

\keywords{editorials, notices --- 
miscellaneous --- catalogs --- surveys}



\section{Introduction}

  The first indications concerning a dual (or multiple) Galactic halo arose from the
 confrontation between the scenario of Eggen, Lynden-Bell, \& Sandage (1962, ELS) and
 that of Searle \& Zinn (1978, SZ).  On one hand, the monolithic collapse, or free-fall, model of  ELS, derived from various orbital-parameter versus ultraviolet-excess diagrams for 221
 "well-observed" dwarf stars, which showed correlations suggesting a nearly
 free-fall collapse. On the other hand, the infall of "protogalactic fragments" proposed by SZ from the differences in the composition found between inner- and outer-halo globular clusters. Several reviews presented the strengths, weaknesses, and similarities of these two scenarios, such as Sandage (1986), Gilmore et al. (1989), and
 Majewski (1993).
 
 
 It was suggested by some authors that these two contrasting views of halo formation were related to differences in the tracers themselves, halo field stars compared to globular clusters, or bias arising from their proper-motion based selection (Mihalas \& Binney 1981; Yoshii 1982; Norris et al. 1985; Chiba \& Beers (2000)). However, later studies using relevant observations discussed the implications and
 importance of combining such ideas in a dual-halo model for the Galaxy (Zinn 1993). For example, evidence for two Galactic halo components was found (M\'arquez \& Schuster (1994), Carollo et al. (2007, 2010), Mar\'in-Franch et al. (2009),
 de Jong et al. (2010), Jofr\'e \& Weiss (2011), Beers et al. (2012)), using $uvby$--$\beta$ photometry of halo field
 stars, globular clusters, and data from the Sloan Digital Sky Survey (SDSS; York et al. 2000) and the sub-program Sloan Extension for Galactic Understanding and Exploration (SEGUE; Yanny et al. 2009) and its extension, SEGUE-2.
 
  Other studies revealed an even more complicated scheme for the Galactic stellar halo, with the discovery of streams, shells, clumps, tidal tails, debris, and the presence of correlated substructures of halo stars (e.g., Koposov et al. 2012; Schlaufman et al. 2009, 2011, 2012; Duffau et al. 2014; Carlberg, Grillmair \& Hetherington 2012; Slater et al. 2014; Carlin et al. 2016), pointing to a more chaotic dual, or even triple, component halo system (see Morrison et al. 2009), extrapolating beyond the ideas of SZ.  Reviews concerning such
 observations and substructures for the stellar halo are given in Helmi (2008), De Lucia
 (2012), Belokurov et al. (2014), and Bernard et al. (2016).



\begin{figure}
	\includegraphics[scale=0.47, trim=20 5 20 15]{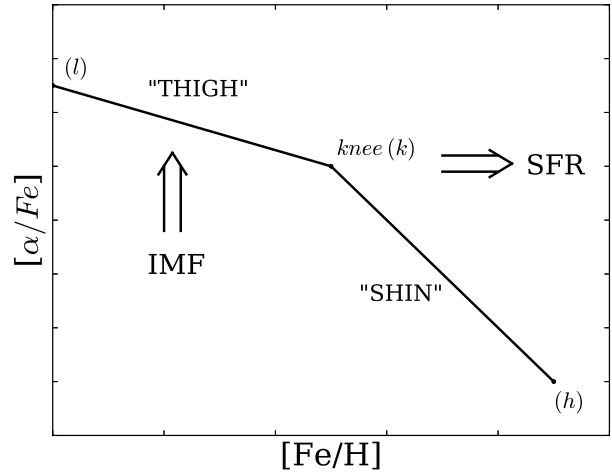}
    \caption{Scheme that we use to identify the chemical trends observed in the [Mg/Fe] vs. [Fe/H] (and other $\alpha$-elements) for our sample (see Figures \ref{fig:mgfe} and \ref{fig:alpha3}). Each line corresponds to a specific stage of the ISM enrichment due to a particular stellar mass contribution, as it is explained in subsection 3.1. The [$\alpha$/Fe] level of the "thigh" depends on the stellar yields and the $M_{up}$ of the IMF. The position of the $knee$ corresponds to a particular [Fe/H] value, which depends on the SFR at early time and the starting time of the bulk of SNIa explosions. The names $(l)$ and $(h)$ refer to the lower and higher [Fe/H] values (as described in the text).}
    \label{fig:esquema}   
\end{figure}

	 Chemistry is considered a valuable tool to sort out the formation processes of the Galaxy. The chemical composition of stellar atmospheres resembles, in most cases, the composition of the interstellar medium (ISM) from which these stars formed. This ISM was chemically enriched by previous stellar populations that contributed their yields of elements (material synthesized by the star and ejected to the ISM) once they reached their last stages of evolution and died. The chemical species synthesized in the stellar interiors depend on the stellar properties, mainly the mass. Thus, the chemical composition measured in presently observed stellar atmospheres provides information about the properties of the previous stellar populations, such as the initial mass function (IMF) or the star-formation rate (SFR) at early times -- see Figure \ref{fig:esquema}. These properties also allow us to constrain the processes that our Galaxy underwent during its early assembly.

Signatures of a dichotomy in the $\alpha$-to-iron ratios in halo stars were detected (Nissen \& Schuster 1997; Fulbright 2002; Gratton et al. 2003; Ishigaki et al. 2013), related in some cases to the distance from the Galactic center (Ishigaki et al. 2010).  Then, in a series of papers, Nissen \& Schuster (2010, 2011 -- hereafter NS10, NS11; and Schuster et al. 2012)
 obtained high-precision ($\pm 0.01$--0.04 dex) relative abundance ratios for 94 dwarf stars
 over the metallicity range $-1.6 <$ [Fe/H] $< -0.4$, with 78 having halo kinematics according
 to the Toomre diagram, plus 16 thick-disk stars. Two groupings were clearly found in
 diagrams such as [Mg/Fe] vs. [Fe/H] or [$\alpha$/Fe] vs. [Fe/H], with 56 high-$\alpha$ halo
 and thick-disk stars falling together, along with 38 low-$\alpha$ halo stars in the
 sample. 
 
  Clear separations between these two halo components were found for the elements Mg,
 Si, Ca, Ti, Na, Ni, Cu, and Zn with respect to iron, and also for [Ba/Y], all as a function
 of [Fe/H] (Ram\'irez et al. 2012 confirmed the same for [O/Fe]).  In Schuster et al. (2012)
 it was shown that the high-$\alpha$ halo stars have ages higher by 2--3 Gyr than the
 low-$\alpha$ ones, and also smaller average values for the orbital parameters  $r_{max},
 z_{max}$, and $e_{max}$.  Again, some concordance with the ideas of ELS plus SZ were found
 (for example, via the $in$ $situ$ and accreted stellar populations of Zolotov et al. 2009, 2010; see also Tissera et al. 2013)


 
 We should note that the distinction between two populations of stars
for high [$\alpha$/Fe] vs. [Fe/H] has also been observed in between
the thick and thin disk (Hayden et al. 2015), and also in other galaxies different
than the Milky Way. Recently, Walcher et al. (2015) demostrated that early
type galaxies present these two populations, associated with older
(alpha-enhanced) and younger (alpha-deficient) populations. Actually,
the star-formation histories of both populations are different, with 
the first one presenting a more sharp one, with a smaller decay time.
This result agree with the strong age-[$\alpha$/Fe] correlation found
in both the Milky Way and other galaxies (e.g., Walcher et al. 2016)

Recent halo studies have benefited from the considerably larger SDSS stellar database with observations at numerous directions in the sky. In particular, the SDSS intermediate-resolution stellar spectra database has allowed the exploration of chemical trends in halo stars over a broad range of distances, up to $\sim$ 50 kpc from the Galactic center, revealing gradients in [Fe/H], [Ca/H] and [Mg/H] (Fern\'andez-Alvar et al. 2015). Yoon et al. (in preparation) argue that the outer-halo gradient continues out to at least 80-100 kpc.

At present, another SDSS program, the APO Galactic Evolution Experiment (APOGEE; Majewski et al. 2015; Nidever et al. 2015) is gathering high-resolution, high signal-to-noise near-IR spectra to map the principal components of the Milky Way. With, eventually, a half million spectra, the APOGEE database is a very valuable sample to check previous findings, and to more completely investigate the chemical properties of stellar populations. Recently, investigations of metal-poor stars in the APOGEE database have showed signatures of the two chemically distinct populations revealed by Nissen \& Schuster (Hawkins et al. 2015; Hayes et al., hereafter Paper I). 

Encouraged by the possibilities of a chemical analysis of the two halo populations discovered in the APOGEE database, we aim to obtain information on the IMF, stellar yields, and star formation history (SFH), or equivalently, SFR vs. time, from the DR13 (Albareti et al. 2016) chemical abundances provided by APOGEE. This paper is organized as follows. The sample selection is discussed in Section 2. Section 3 describes the split of the sample into two populations, the derivation of the corresponding chemical trends, and the theoretical model from which we infer properties for each population. In Section 4 we relate our main results, and we discuss them in Section 5. Section 6 summarizes our conclusions.

\section{Sample} \label{sec:sample}

APOGEE is an SDSS program (Eisenstein et al. 2015; Blanton et al. 2017) conceived to explore the structure of the Milky Way. The first APOGEE phase was in SDSS-III and collected data between 2011 and July 2014, obtaining high-resolution (R$\sim22,500$) spectra with a typical signal-to-noise $\geq 100$ using a multiobject infrared spectrograph coupled to the 2.5 meter SDSS telescope at Apache Point Observatory (Gunn et al. 2006, Wilson et al. 2010). The targets map the Galactic disk, bulge, and halo (Zasowski et al. 2013). More than 143,000 objects were observed as part of that program. 

The APOGEE Stellar Parameters and Abundances pipeline (ASPCAP) was developed to obtain stellar atmospheric parameters and chemical abundances from the $H$-band (1.5-1.7 $\mu$m), the spectral range covered by the APOGEE spectrograph. The methodology is based on the comparison with synthetic spectra in an N-dimensional parameter space, looking for the best fit with observations (more details in Garc\'ia-P\'erez et al. 2016). Abundances with accuracies $\sim 0.1$ dex have been derived, and radial velocities have been determined with accuracies of $\sim 0.1$ km/s (Holtzman et al. 2015). DR12 (Alam et al. 2015) was the final SDSS-III data release. The thirteenth data release (DR13; Albareti et al. 2016) provides the final products of a re-analysis, after including several improvements to the pipeline. Chemical abundances of up to 26 chemical species are available for some stars, including the $\alpha$-elements: O, Mg, S, Si, Ca and Ti.

From this database we want to draw a sample of halo stars. These objects clearly exhibit different kinematics from disk stars; objects with large heliocentric radial velocities have a high probability to belong to this Galactic component. 

In addition, the $l-GRV/cos(b)$ space (GRV is the radial velocity $v_{rad}$ corrected for solar motion\footnote{We adopt the solar Galactocentric velocities $U_{\odot}=11.1$ km/s, $V_{\odot}=12.24$ km/s and $W_{\odot}=7.25$ km/s (Brunthaler et al. 2011)}, and $l$ and $b$ the galactic longitude and latitude) can be used to isolate halo stars from disk stars. As performed by Hawkins et al. (2015) for the same purpose, we exclude from our sample those stars following a sinusoid of amplitude corresponding to the rotational velocity of the disk in the solar circle (220 km/s; Sch\"onrich 2012) and a dispersion more than three times the dispersion of the same curve defined by disk stars. 

This is the sinusoid expected to be drawn by objects rotating in the Galactic plane. Halo stars occupy randomly the GRV/cos(b) vs. l space and, consequently, this selection criteria excludes not only disk stars but also stars belonging to the Galactic halo. However, we prefer to select only those objects with the highest confidence to be halo stars, even if our selection criteria is quite restrictive. 

This selection to exclude disk stars works best in the case of objects at $|b| < 60 \degr$ (Majewski et al. 2012). Therefore, we measured the dispersion for stars at $|b| < 60\degr$, having $\rm[Fe/H] > 0.0$, which we expect to be dominated by disk stars. Thus, to explore stars with halo kinematics we select objects with $v_{rad}$ $> 180$ km/s and/or stars at$|b| < 60 \degr$ with an absolute values of GRV/cos(b) more than three times larger than that measured with disk stars in bins of $20\degr$ in $l$.

The key to identifying different stellar populations by their chemistry is the accuracy and precision with which their chemical abundances are measured (Lindegren \& Feltzing 2013). Both are also needed to infer parameters of the SFH from their chemical abundance trends. The random abundance uncertainties in the ASPCAP analysis vary as a function of $T_{\rm eff}$ and [Fe/H], as illustrated in Figure 2 of Bertr\'an de Lis et al. (2016). They evaluated the [O/Fe] uncertainty as a function of these parameters by measuring the scatter observed in clusters. In light of these results, we select stars in the $T_{\rm eff}$ range at which the precision in [O/Fe] is the highest for the metallicity range covered by our halo sample (-2.5 $<$ [Fe/H] $<$ +0.5). We choose the interval 4000 $< T_{\rm eff} <$ 4500 K, where the [O/Fe] uncertainties are $\sigma$[O/Fe] $\leq0.02$ dex, for -0.6 $<$ [Fe/H] $<$ +0.2, and increasing at lower metallicities. The empirical uncertainties calculated by ASPCAP are, on average, $\delta = 0.05$ dex for [O/Fe] and [Mg/Fe], with a standard deviation $\sigma = 0.03$ dex and 0.02 dex, respectively, and $\delta = 0.04$ dex and 0.07 dex with $\sigma = $0.02 dex and 0.06 dex, for [Si/Fe] and [Ca/Fe], respectively. 

We exclude stars with distances (adopted using the techniques of the Brazilian Participation Group -- BPG; Santiago et al. 2016) from the Galactic center $r \leq 4$ kpc, in order to avoid bulge stars. We know that this selection cut may exclude also some halo stars located at this range of distances. However we want to avoid any possible bulge contamination due to our goal in characterizing only the halo component of our Galaxy.

Finally, we reject objects with ASPCAP flags indicating possibly poor estimates. We also reject targets in globular clusters in our Galaxy and in Andromeda -- these show chemical abundances that strongly deviate from the chemical trends of field stars (Meszaros et al. 2015). Our sample is comprised by field stars. We do not expect to have included objects from dwarf spheroidal galaxies, because APOGEE only purposely targeted in DR13 the spheroidal Sagittarius. Most of their stars are cool M giants with $T_{\rm eff} < 4000$ K, and we rejected them by our $T_{\rm eff}$ selection criteria.

Each star of our final sample is assigned to belong to the Galactic halo merely based on its $v_{rad}$. A more robust attribution will be possible with Gaia parallaxes and proper motions very soon. 

Within the $\alpha$-elements derived by ASPCAP, S and Ti are less reliable. [Ti/Fe] derived from both neutral and ionized atomic lines shows a large dispersion, and differing trends in the [Ti/Fe] vs. [Fe/H] space. In their evaluation of the ASPCAP products, Holtzman et al. (2015) warned about the reliability for this element's abundances. They found no trend of [Ti/Fe] with [Fe/H], which led them to suspect that some systematic error affected their measurement. [S/Fe] also exhibits a large dispersion. At low metallicities, the S I lines from which ASPCAP derives [S/Fe] abundances become weak and comparable to the noise level. The measured [S/Fe] abundances lead to enhanced values and are likely to be unreliable. 


\begin{table*}
\renewcommand{\thetable}{\arabic{table}}
\centering
\caption{Halo stars selected within the DR13 APOGEE database, with the stellar atmospheric parameters and chemical abundances determined by ASPCAP and used in this work.}\label{tab:data} \footnote{Full table available at the CDS.}
\begin{tabular}{rccccccccc}
\tablewidth{0pt}
\hline
\hline
 2MASS ID &  Population & $T_{\rm eff}$ & $\log g$ & $\rm v_{rad}$ & $\rm [Fe/H]$ & $\rm [O/Fe]$ & $\rm [Mg/Fe]$ & $\rm [Si/Fe]$ & $\rm [Ca/Fe]$ \\
 \hline
 \decimals

      2M13590274+0118564  & HMg &     4486   &         1.0 &    259.790   &       -1.95  &         0.53   &        0.39   &        0.27    &       0.29 \\
            2M23161405+1257322 &  HMg  &    4435     &       1.1  &  -179.804   &       -1.86    &       0.30    &       0.49      &     0.66    &       0.16  \\
            2M10374221-1042328  &   HMg  &    4346    &        1.2  &   181.805    &      -1.34     &      0.43     &      0.31       &    0.12    &       0.43  \\
            2M11002833-1044050  &  HMg    &  4483      &      1.1  &   337.768     &      -1.28    &   -    &       0.31     &      0.40    &       0.38  \\
            2M11422622-1409451 &  HMg  &    4323    &        1.2  &   258.096     &     -1.20    &       0.36    &       0.24    &       0.20     &      0.33  \\
\hline\hline
\end{tabular}
\end{table*}

\section{Methodology} \label{sec:methodology}

\begin{figure}
	\includegraphics[scale=0.46,trim=0.25cm 6cm 2.6cm 7.5cm, clip=true]{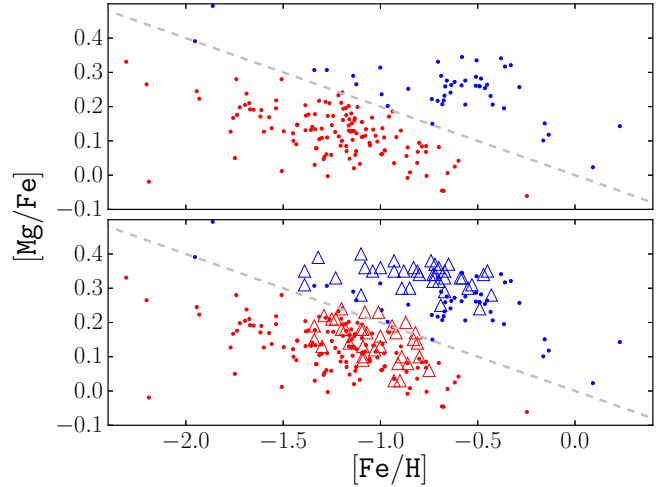}
    \caption{[Mg/Fe] as a function of [Fe/H] for our sample in both panels, with the NS10 sample overplot in the bottom panel. APOGEE data are shown as dots. Triangles represent the [Mg/Fe] values from NS10. The black dashed line separates the sample in two populations as in Paper I, along [Mg/Fe] $= -0.2$[Fe/H]. We extrapolate the separation to lower [Fe/H]. The High-Mg population is shown with blue symbols, while the Low-Mg is shown with red symbols.}
    \label{fig:mgfe}   
\end{figure}

Our final sample comprises 175 stars. The top panel of Figure \ref{fig:mgfe} shows the comparison between [Mg/Fe] as a function of [Fe/H]. Two different chemical trends are clearly distinguishable. We split the two stellar populations (High-Mg and Low-Mg) following the same classification derived from the statistical analysis presented in Paper I from a larger sample, i.e., along [Mg/Fe] $= -0.2$[Fe/H]. In the bottom panel we overplot the NS10 sample of halo stars with their [Mg/Fe] and [Fe/H] measurements. Their sample abundance trends follow very closely our results from APOGEE data.  



\begin{figure}
	\includegraphics[scale=0.42, trim= 0cm 5cm 0cm 6cm, clip=true]{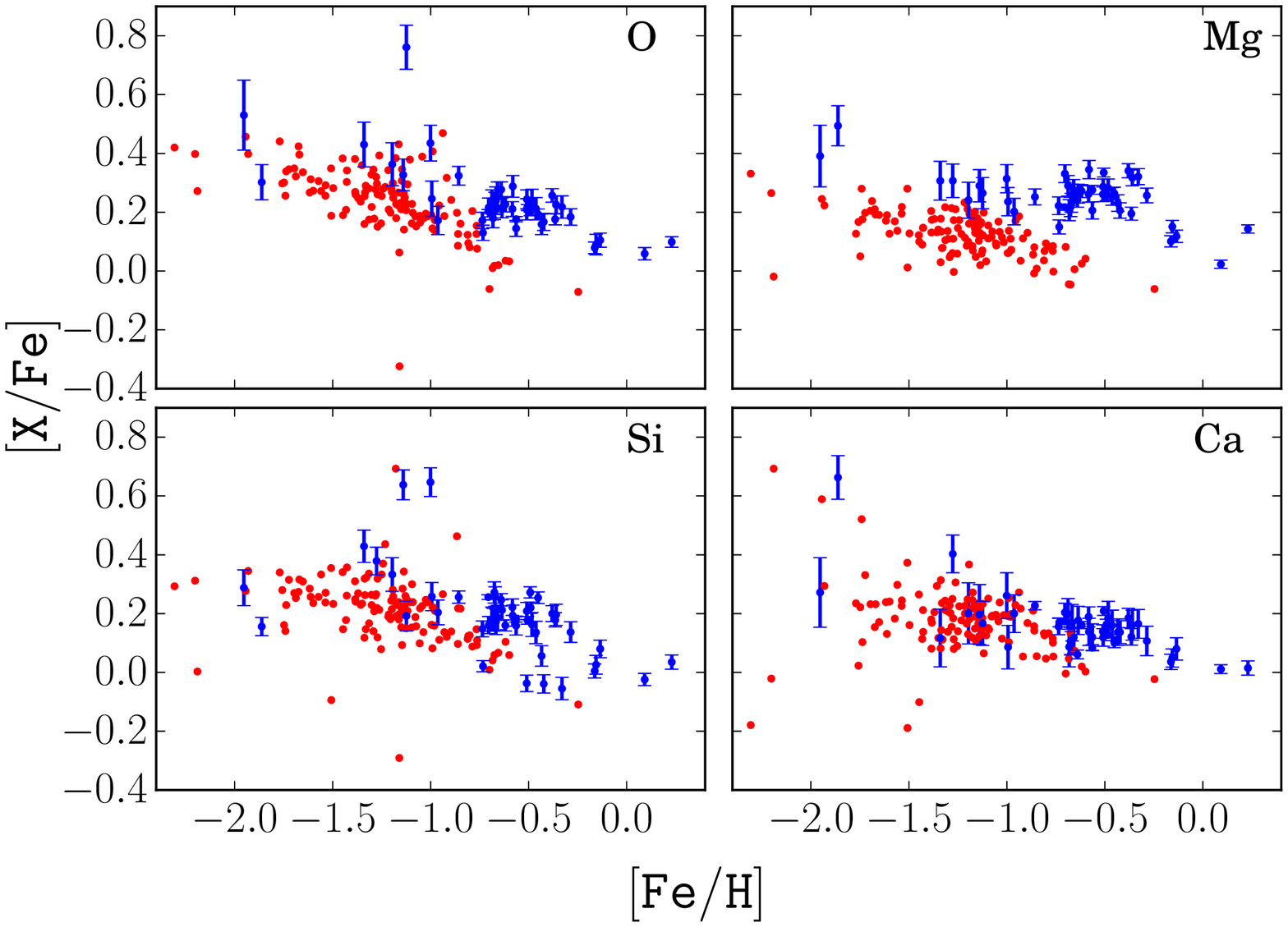}
  	\includegraphics[scale=0.42, trim= 0cm 5cm 0cm 7.5cm, clip=true]{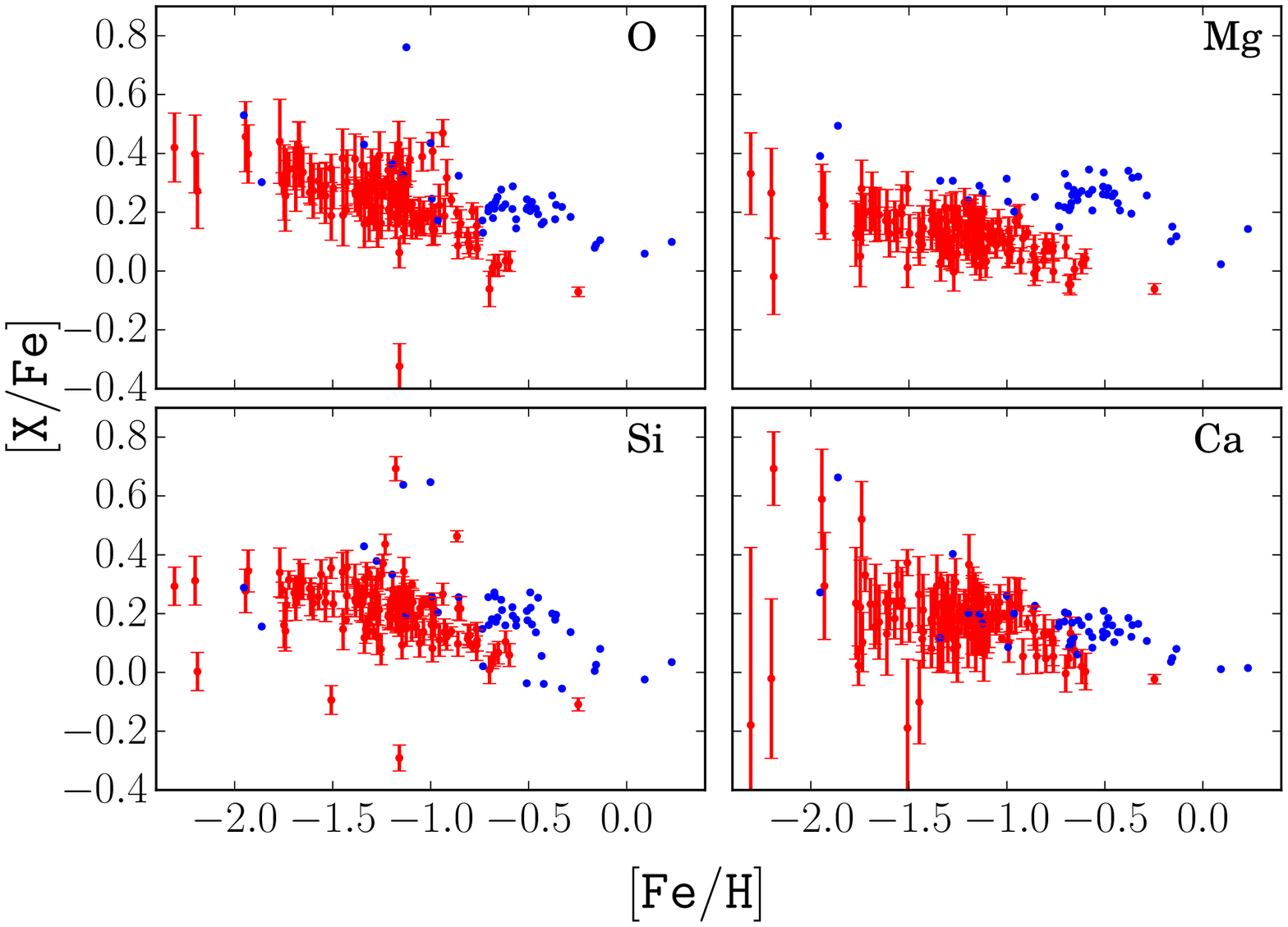}
    \caption{[X/Fe] as a function of [Fe/H] for the reliably ASPCAP measured chemical elements, O, Mg, Si and Ca, with their associated errors for the \halpha{} (top panels) and for the \lalpha{} (bottom panels) populations.}
    \label{fig:alpha}   
\end{figure}



We investigate these two populations in the [X/Fe] vs. [Fe/H] space for the other APOGEE-reliable $\alpha$-chemical species determined by ASPCAP. The chemical abundances and their errors are displayed in Figure \ref{fig:alpha} (for the High-Mg population, upper panels, and for the Low-Mg population, lower panels). In order to better visualize the chemical trends and the differences between the two populations, we calculate the weighted mean [X/Fe] and its statistical error in [Fe/H] bins of 0.1 dex, with a minimum of five objects per bin. 

\begin{figure*}
	\includegraphics[scale=0.75, trim= -.75cm 6cm 2cm 7cm, clip=true]{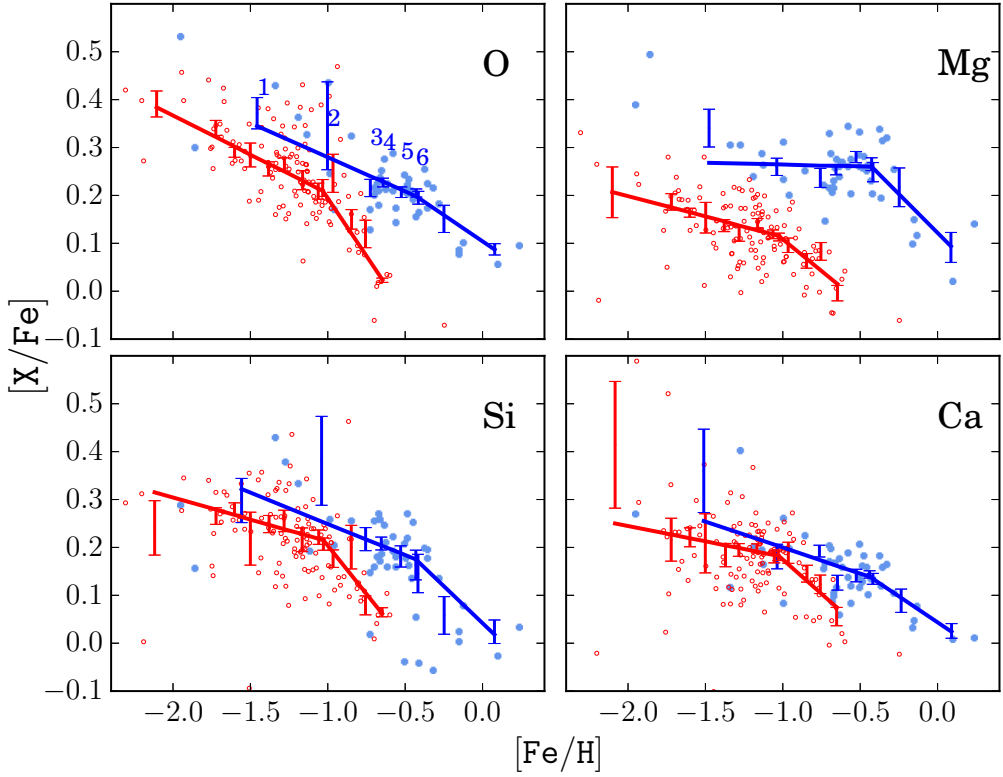}
    \caption{[X/Fe] as a function of [Fe/H] for the $\alpha$-elements O, Mg, Si and Ca in each panel. The populations are color-coded as in Figure \ref{fig:mgfe}. The weighted mean [X/Fe], calculated in bins of 0.1 dex in [Fe/H] (considering a minimum of 5 objects per bin), with their corresponding errors, are overplotted in blue and red for the \halpha{} and \lalpha{} populations, respectively.
    Linear fits to the weighted [X/Fe] means are overlapping (blue and red solid lines). The six more metal-poor values resulting from the fit and use to infer the IMF $M_{up}$ are indicated with numbers in the left top panel corresponding to [O/Fe] vs. [Fe/H].} 
    \label{fig:alpha3}   
\end{figure*}

Figure \ref{fig:alpha3} shows the [X/Fe] vs. [Fe/H] for each population.
The High-Mg population shows the largest enhancement of all the $\alpha$-elements considered here. This broad separation is the reason we used this element as the primary discriminator of halo populations in Paper I. The enhancement level diminishes with  O, Si, and Ca. For this reason, and to be consistent with the nomenclature in the first paper of this series, we refer to these populations as High-Mg (HMg) and Low-Mg (LMg).

The [X/Fe] vs. [Fe/H] trends (see Figure \ref{fig:alpha3})
can be divided into two parts, as we depict in Figure \ref{fig:esquema}:

\begin{enumerate}

\item The "thigh": This corresponds to the semi-$plateau$ located between the lowest metallicity ($\rm[Fe/H]_{l}  \sim -1.9 $ and $\sim -1.4$ for the \lalpha{} and \halpha{} populations, respectively) and the $knee$ ($\rm[Fe/H]_{k} \sim -1.0$ and $\sim -0.4$ for the \lalpha{} and \halpha{} populations, respectively). 
The $\rm[Fe/H]_{k}$ is the metallicity at which the downward slope becomes steepest.


\item The "shin": Located between the metallicity of the $knee$ and the largest metallicity, [Fe/H]$_{h}$.  

 

\end{enumerate}





The chemical trends observed for the \halpha{} and \lalpha{} sub-samples in each one of these metallicity ranges are the result of a different chemical and star formation history for each stellar population, as explained below.


We use the calculated weighted means, choosing bins of different sizes to determine the mean [Fe/H] at which the slope of the trend changes, corresponding to the $knee$ of the population. We identify the \halpha{} $knee$ at [Fe/H] $\sim-0.4$, and the \lalpha{} $knee$ at [Fe/H] $\sim-1$.

It is important to notice that not all the elemental abundances could be measured for every star. 
In some cases, it was not possible to determine the abundance of a particular element reliably due to the quality of the observations. 
For this reason, the number of stars in each sample slightly varies from one element to another, as well as the particular objects from which the means are calculated. 
This implies that the mean [Fe/H]$_{l}$, [Fe/H]$_{k}$, and [Fe/H]$_{h}$ for each population are slightly different, depending on the chemical element that we consider. 
Table \ref{tab:means} shows the resulting various $\rm<[Fe/H]>$ for each $\alpha$-element.
Due to the low number of stars, the weighted means have large errors, in particular at low metallicities, and they do not describe smooth chemical trends. For this reason, we perform a linear fit to the weighted [X/Fe] means in the "thigh" and in the "shin" metallicity ranges; see Figure \ref{fig:alpha3}.

\begin{table*}
\renewcommand{\thetable}{\arabic{table}}
\centering
\caption{Mean [Fe/H] and [X/Fe] derived from the two stellar populations at the key [Fe/H] values.}\label{tab:means}
\begin{tabular}{cccccccc}
\tablewidth{0pt}
\hline
\hline
 Population & element &  $\rm [Fe/H]_{l}$ & $\rm [X/Fe]_{l}$ & $\rm [Fe/H]_{k}$ & $\rm [X/Fe]_{k}$ & $\rm [Fe/H]_{h}$ & $\rm [X/Fe]_{h}$ \\
 \hline
 \decimals
 \halpha{} & O & $-1.46\pm0.15$ &   $0.37\pm0.03$ & $-0.43\pm0.02$ &  $0.20\pm0.01$ &  $0.08\pm0.09$ & $0.09\pm0.01$  \\
   & Mg &  $-1.48\pm0.13$ & $0.34\pm0.04$ & $-0.43\pm0.02$ & $0.26\pm0.01$ & $0.09\pm0.08$ & $0.09\pm0.03$  \\
   & Si & $-1.56\pm0.14$ &  $0.30\pm0.05$ & $-0.43\pm0.02$ & $0.16\pm0.03$ & $0.08\pm0.08$ & $0.02\pm0.02$ \\
   & Ca & $-1.51\pm0.14$ & $0.36\pm0.09$ & $-0.44\pm0.02$ & $0.14\pm0.01$ & $0.09\pm0.07$ & $0.03\pm0.01$  \\
\hline
 \lalpha{} & O & $-2.11\pm0.07$ & $0.39\pm0.03$ &  $-1.03\pm0.01$ & $0.21\pm0.02$ & $-0.65\pm0.01$ & $0.02\pm0.00$ \\
   & Mg & $-2.10\pm0.07$ & $0.21\pm0.05$ & $-1.02\pm0.01$ & $0.11\pm0.01$ & $-0.65\pm0.01$ & $-0.00\pm0.02$ \\
   & Si & $-2.12\pm0.07$ & $0.24\pm0.06$ & $-1.03\pm0.01$ & $0.21\pm0.01$ & $-0.65\pm0.01$ & $0.06\pm0.01$ \\
   & Ca & $-2.08\pm0.06$ & $0.41\pm0.13$ & $-1.03\pm0.01$ & $0.18\pm0.01$ & $-0.65\pm0.01$ & $0.06\pm0.02$ \\
\hline\hline
\end{tabular}
\end{table*} 



\subsection{Chemical-Evolution Model}

As mentioned in the Introduction, we aim to obtain basic chemical histories for the \halpha{} and \lalpha{} populations.
In particular, we try inferring: the upper mass ($M_{up}$) of the IMF,  the integrated yields for massive stars ($Y$),  the fractions of Type II supernovae (SNII) and Type Ia supernovae (SNIa) ($f_{SNII}$ and $f_{SNIa}$, respectively) in each simple stellar population, and
the efficiency ($\nu$) of the SFR and duration ($t_h$) of the SFH.

In order to obtain general properties for each population, our simple chemical-evolution models are built based on the following assumptions:

\begin{enumerate}

\item The \halpha{} and \lalpha{} are two independent populations. 
Each population evolves in its own way, according to the trend described by the mean [X/Fe] values for O, Mg, Si, and Ca vs. [Fe/H] (See Figure \ref{fig:alpha3}). 


\item We embrace the semi-instantaneous recycling approximation. 
In this approximation, after each burst of star formation, all massive stars explode as SNII, instantaneously enriching the interstellar medium (ISM). SNIa explode with a delay of about 1 Gyr after their progenitors are formed. For a similar prescription, see Franco \& Carigi (2008) and Hern\'andez-Mart\'inez et al (2011).

\item A closed-box model that evolves from initial primordial gas.
 We assume one zone per population, and a continuous SFH.


\item Each [X/Fe] vs. [Fe/H] range represents a different evolutionary stage:
\begin{enumerate}
\item During the "thigh",  only SNII contribute to the ISM enrichment.
\item During the "shin", SNII and SNIa pollute the ISM. 
These SNII behave similarly to the SNII in the "thigh".
\end{enumerate}

\end{enumerate}

Based on assumptions 2 and 3, the chemical abundance by mass ($X$\footnote{We differentiate the chemical abundance by mass from the chemical abundance by number of an element indicating the former with an italicized font.}) of an element in the interstellar medium evolves between any two times, $t_{1}$ and $t_{2}$ ($> t_{1}$)

\begin{equation}
\Delta X = X(t_2) - X(t_1) = -<Y_X>_{2-1} \log {{\frac{\mu(t_2)}{\mu(t_1)}}}, 
\end{equation}

\noindent
where
$Y_X$ is the synthesized mass fraction of element $X$ ejected by dying stars,
$<Y_X>_{2-1}$ represents the $Z$-average integrated yields between  $Z(t_1)$ and $Z(t_2)$ (i.e., $Z(t_j)=0.02\times10^{[Fe/H]_j}$), $\mu(t)  = M_{gas}(t)/M_{gas}(0)$ is the gas consumption, $M_{gas}$ represents the gas mass, and $M_{gas}(0)$ is the initial gas mass (see Avila-Vergara et al. 2016).

For computing $<Y_X>$,  we consider theoretical $Z$-dependent yields for SNII by Kobayashi et al. (2006) and for pre-SN by the Geneva group (see Robles-Valdez et al. 2013). We integrate these yields in mass over a Kroupa-Tout-Gilmore IMF (Kroupa et al. 1993). The integrated yields are calculated between $0.1 M_{\odot}$ and $M_{up}$,  where $M_{up} = $ 10 to 40 in steps of 5  $M_{\odot}$. 
For SNIa we assume the $Z$-independent SNIa yields by Iwamoto et al. (1999).

Applying Eq. 1 to $\alpha$ and $Fe$, we derive: 

\begin{equation}
\frac{\Delta X(t_2)}{\Delta X(t_1)} = \frac{X(t_2)-X(t_1)}{Fe(t_2)-Fe(t_1)} =\frac{<Y_X>_{2-1}}{<Y_{Fe}>_{2-1}} 			
\end{equation}

\noindent
or equivalently,
\begin{equation}
\frac{\Delta X/H}{\Delta Fe/H} = \frac{X_{2}/H-X_{1}/H}{Fe_{2}/H-Fe_{1}/H} = \frac{<Y_X>_{2-1}}{<Y_{Fe}>_{2-1}}    
\end{equation}

Eq. 2 relates the abundance ratios (derived by ASPCAP from the observations) with the integrated yields (from theoretical yields and the IMF). For obtaining the best $M_{up}$ values that reproduce the data we apply Eq. 2 on the "thigh", because during this range only SNII enrich the ISM.

Based on the data, we can obtain X/H (the fraction by number) from [X/Fe]$ - $[Fe/H], taking into account the solar abundances by Asplund et al. (2005). These are the solar chemical abundances considered by ASPCAP in the generation of the synthetic spectra of the grids used to determined the elemental abundances from APOGEE spectra. We calculate $X/H$ and $Fe/H$ ratios from the values derived by the linear fit over the weighted [X/Fe] means shown in Figure \ref{fig:alpha3}. Then, we transform the X/H value by number to $X/H$ by mass. 


\subsubsection{The "thigh"}

\begin{figure}
\includegraphics[scale=0.43, trim=0 0 0 -0.5cm, clip=true]{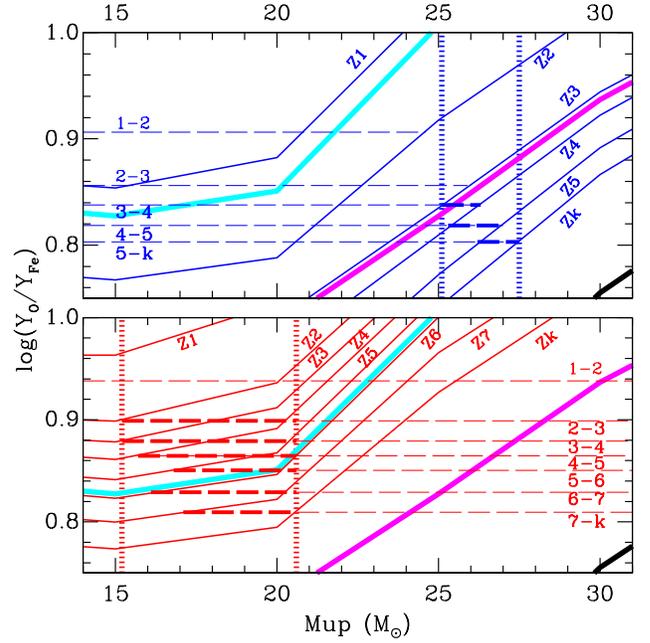}
\caption{Comparison of the oxygen yield to iron yield ratio ($Y_{O}/Y_{Fe}$) with the $\Delta O/\Delta Fe$ values (see Eq. 3) for the \halpha{} population (upper panel) and the \lalpha{} population (lower panel).
Inclined continuous lines: $Y_{O}/Y_{Fe}$, results of integrating IMF-weighted-yields from 8 $M_{\odot}$ to $M_{up}$ (10, 15, 20, 25 and 30 $M_{\odot}$) 
for different initial stellar metallicities.
Inclined thick lines: $Y_{O}/Y_{Fe}$ for the initial metallicities (0.001, 0.004, and 0.02; cyan, magenta and black) considered by Kobayashi et al (2006).
Inclined thin lines: $Y_{O}/Y_{Fe}$ for the metallicities ($Z_{j}$) corresponding to the [Fe/H]-means of the "thigh" (k values).
Horizontal dashed lines: $\Delta O/\Delta Fe$, obtained from the linear fits (blue and red solid lines of Figure \ref{fig:alpha3}) and their consecutive [O/Fe] and [Fe/H] means in the "thigh".
Horizontal thin lines: $\Delta O/\Delta Fe$ between two consecutive values of the linear fits (1-2,2-3,…).
Horizontal thick lines: Most reliable pair of these values and the associated $M_{up}$. 
Vertical dotted lines: lower and upper $M_{up}$, inferred from the intersection of $Y_{O}/Y_{Fe}$ and the reliable $\Delta O/\Delta Fe$.}
\label{Mup}
\end{figure}
We infer the $M_{up}$ of the IMF for each population from the "thigh". We use O abundances because, in the literature, chemical evolution models cannot reproduce the [Mg/Fe]-[Fe/H] trend shown by stars of the solar vicinity.
For example, according to Romano et al. (2010), Mg in halo stars is reproduced when models considered yields by Kobayashi et al. (2006) for supernovae and hypernovae, but with this yield combination Mg in disk stars does not fit, mainly in thick-disk stars.

Figure \ref{Mup} shows the theoretical $\log(Y_{O}/Y_{Fe})$ vs. $M_{up}$
for stellar initial metallicities equal to 0.001, 0.004, and 0.02 (cyan, magenta and black lines).
We add the yield ratios interpolated for the metallicities ($Z_{j}$) that corresponds to the [Fe/H] means in Figure \ref{fig:alpha3} lower than the $knee$, $\rm[Fe/H] \leq [Fe/H]_k$.

We depict $log(\frac{X_{j+1}/H-X_{j}/H}{Fe_{j+1}/H-Fe_{j}/H})$ using horizontal lines.
For convenience, we focus on the $X/Fe$ that show lower errors. From the intersection between the theoretical yields and the measured abundances we infer the $M_{up}$.
The inferred $M_{up}$ ranges are located between vertical dotted lines (for the high- and low- $\alpha$ populations, upper and bottom panels, respectively). 
Based on the mean $ M_{up}$ values corresponding to the inferred $M_{up}$ ranges, we compute the fraction of SNII ($\int_{8}^{M_{up}}$IMF($m$)$dm$) for each simple stellar population
and the integrated yields between $8 M_{\odot}$ and $M_{up}$.



Considering that theoretical Fe yields for massive stars, $<Y_{Fe}^{theo,II}>$, are well-computed and do not require correction, 
we employ Eq. 2 to obtain the empirical yields, $<Y_X^{emp,II}>$, needed to reproduce the observed [X/Fe] vs. [Fe/H] between the fourth mean point and the $knee$, the most reliable range for the "thigh". 

\begin{equation}
<Y_X^{emp,II}> = <Y_{Fe}^{theo,II}>  \frac{X_{k}/H-X_{4}/H}{Fe_{k}/H-Fe_{4}/H} 
\end{equation}

\begin{figure}
	\includegraphics[scale=0.48, trim= 0.8cm 0cm 1cm 1cm, clip=true]{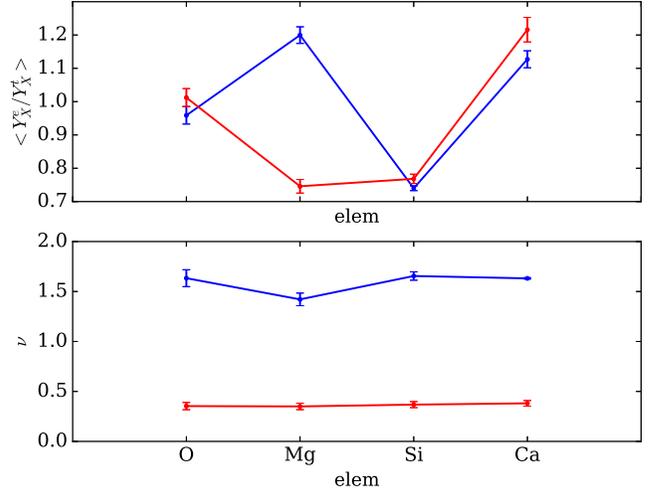}
    \caption{Upper panel: Correction factors, the empirical yield to theoretical yield ratio,$<Y_X^{emp,II}>/<Y_X^{theo,II}>$, inferred to reproduce the observed [X/Fe] values in the "thigh", for the $\alpha$-elements considered in this study and for both halo populations in the APOGEE sample. Bottom panel: Efficiency of the SFR, $\nu$. The error bars show the values derived from the limits of the $M_{up}$ range inferred of each population.}
    \label{fig:1k}   
\end{figure}

Figure \ref{fig:1k} shows the correction factors, $<Y_X^{emp,II}>/<Y_X^{theo,II}>$, for each of the $\alpha$-elements and populations, which are close to unity. 
Error bars are calculated from the minimum and maximum $M_{up}$ (see Table 2) assumed in the computation of $<Y_{Fe}^{theo,II}>$.

Oxygen presents the largest errors, because O yields are the most stellar-Z and mass-dependent ones among the four APOGEE-reliable $\alpha$-elements.
The Mg correction factor is the most different between the populations, due to the high Mg enhancement shown by the \halpha{} population.


\subsubsection{The "shin"}

Again, we apply Eq. 2, but now for the "shin" (between the $knee$ and the $\rm[Fe/H]_{h}$), taking into account the same correction factors for integrated yields of massive stars 
($<Y_X^{emp,II}>$)  obtained in the "thigh" range.
As previously mentioned, SNII and SNIa contribute during the "shin" to the ISM enrichment in alpha and Fe elements. 
Therefore $Z$-average integrated yields ($<Y_X>_{k-h}$) between $Z(t_{k})$ and $Z(t_{h})$ are:

\begin{equation}
<Y_X>_{k-h}= <Y^{emp,II}>_{k-h} + f_{SNIa} \times y_{X}^{Ia},
\end{equation}

\noindent
where $y_X^{Ia}$ is the SNIa yield for a specific element, and $f_{SNIa}$ is the fraction of SNIa that contributed during the "shin".
In this range,  the contribution of low- and intermediate-mass stars are not considered, either because these stars do not produce the $\alpha$-elements considered in this paper, or their yields are negligible compare with the SNII and SNIa yields. 

\begin{figure}
	\includegraphics[scale=0.48,trim= 0.5cm 0cm 1cm 1cm, clip=true]{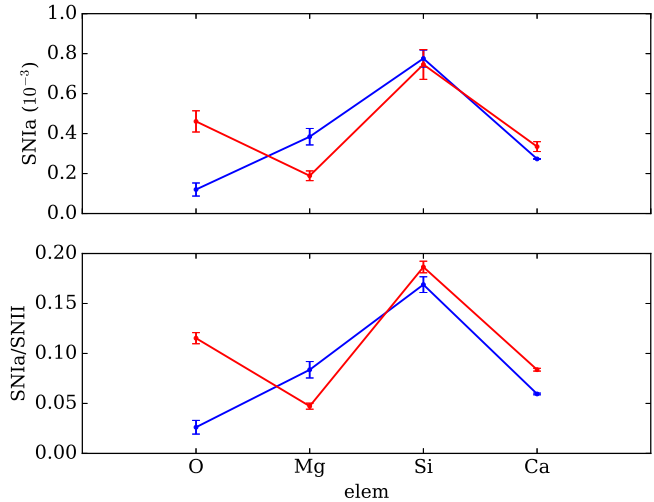}
    \caption{Parameters inferred from the "shin" (between $[Fe/H]_{knee}$ and $[Fe/H]_{h}$). Top panel: Fraction of SNIa that explode per each simple stellar populations of 1$M_{\odot}$. Bottom panel: Fraction of SNIa relative to the fraction of SNII formed in each simple stellar population. The error bars show the values derived from the limits of the $M_{up}$ range inferred of each population.}
    \label{fig:ka}   
\end{figure}

\begin{figure}
	\includegraphics[angle=90,scale=0.37, trim= 0.5cm -1cm -1cm 0]{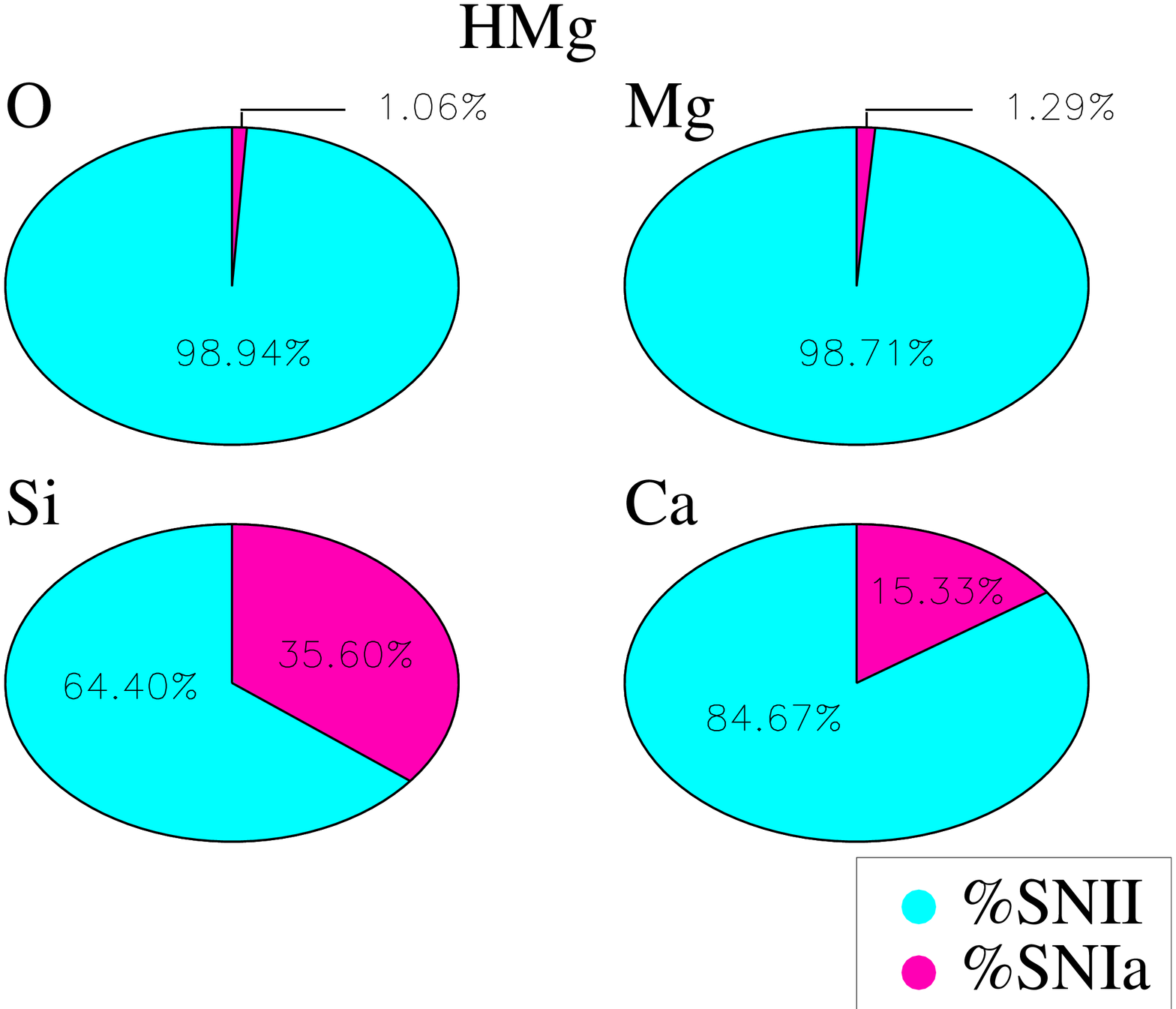}
	\includegraphics[angle=90,scale=0.37, trim= 0.5cm -1cm -1cm 0]{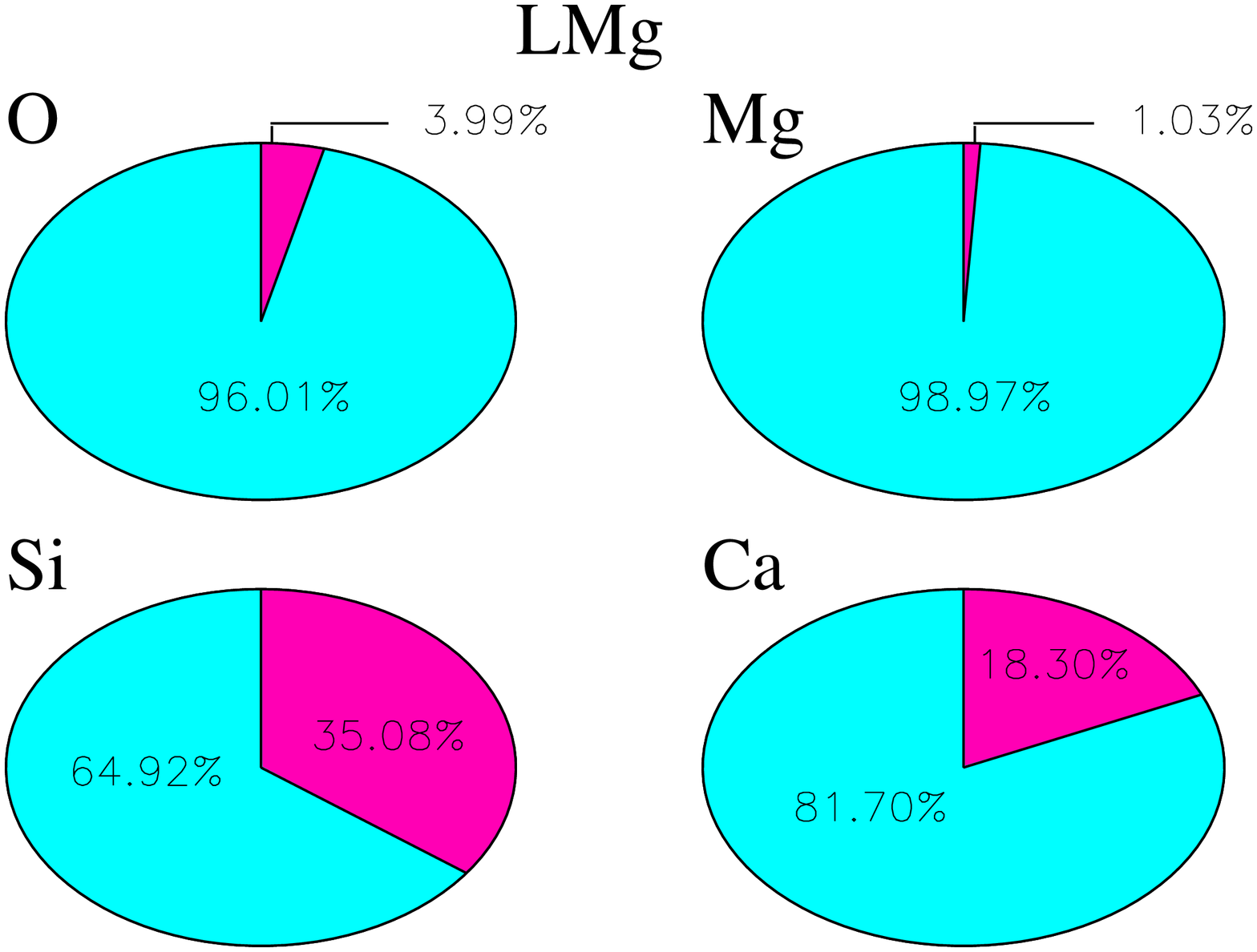}
    \caption{Percentage contribution by SNII (cyan) and SNIa (magenta) to the $\alpha$-enrichment, during the "shin", for the \halpha{} (upper ellipsoids) and \lalpha{} (bottom ellipsoids) populations.}
    \label{fig:ka2}   
\end{figure}

Substituting $<Y_X>_{k-h}$ and $<Y_{Fe}>_{k-h}$ in Eq. 3, we obtain $f_{SNIa}$.
Figure \ref{fig:ka} shows the $f_{SNIa}$ values (upper panel) and $\frac{f_{SNIa}}{f_{SNII}}$ for each $\alpha$-element and population. Figure \ref{fig:ka2} shows the percentage contribution of SNIa and SNII to the $\alpha$-element enrichment.

\hfill \break

Finally, this basic model also allows us to estimate the efficiency of the SFR and 
 the times when $M/H_k$ and $M/H_h$ occur.
 For that, in Eq. 1 we assume that:
 \begin{enumerate}
\item The SFR is proportional to the $M_{gas}$, with efficiency $\nu$, $SFR(t) = \nu Mgas(t)$
\item  $\nu$ is constant during the entire evolution
\item  $t_k =$ delay-time for SNIa $= 1$ Gyr, and
 \item $X(t_j) \sim X_j/H \times H_\odot$, where  $H_\odot=0.7392 $ (Asplund et al. 2005).
  \end{enumerate}
  
 Therefore, 
\begin{equation}
M_{gas}(t) = M_{gas}(0) e^{-\nu(1-R)t},
\end{equation}

\noindent
where $R$ is the fraction of the mass ejected into the ISM by the dying stars.
 
 From $t=0$ until $t_k$, a left-infinite "thigh", we find that
 \begin{equation}
X(t_k) =  <Y_X> \nu(1-R)t_k,
\end{equation}
\noindent
 and we obtain $\nu$.
The bottom panel of Figure \ref{fig:1k} shows the $\nu$ values for $\alpha$-elements and each population.

Focusing on the "shin", we compute $t_h$ from
 \begin{equation}
X(t_h)-X(t_k)  =  <Y_X>_{k-h} \nu(1-R)(t_h-t_k).
\end{equation}


Finally, we calculate the SFR for each population:
\begin{equation}
 SFR(t) = \nu M_{gas}(0) e^{-\nu(1-R)t}.
\end{equation}

\begin{figure*}
	\centering
	\includegraphics[angle=270, scale=0.7,trim= 3.7cm 0cm 0.5cm 0cm, clip=true]{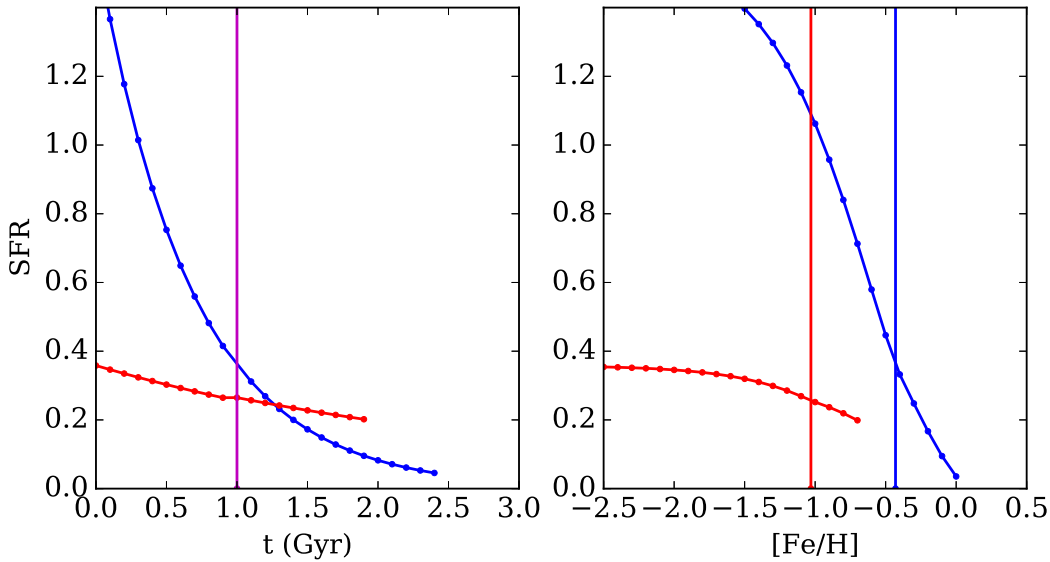}
    \caption{The SFR as a function of time (left panel) and [Fe/H] (right panel). Vertical lines represent the time and [Fe/H] for the $knee$ for each stellar population. The figure is color coded as in Figure 4.}
    \label{fig:sfr}   
\end{figure*}

We compute R using stellar ejecta by Kobayashi et al. (2006) and Karakas et al. (2010).
Before the $knee$, only massive stars contribute to the ISM, and $R=0.061$.
After the $knee$, massive and intermediate-mass stars die, and $R= 0.158$.
We derive the SFR($t$) assuming $t_k = 1$ Gyr and $M_{gas}(0)= 1 M_\odot$. 
These choices are motivated to easily obtain the SFHs for other $t_k$ and $M_{gas}(0)$ values. Figure \ref{fig:sfr} exhibits the resulting SFR($t$) for the \halpha{} and \lalpha{}, in blue and red, respectively. We show the SFH for each population considering
$\nu$ obtained from Fe, due to the familiar time-metallicity relation.

Table \ref{times} presents the times when the final enrichment occurs for the analyzed elements for each population.

\section{Results}

\subsection{Chemical trends}

The resulting trends for both populations are characterized by [X/Fe] decreasing with [Fe/H]. From the weighted [X/Fe] means we identify the $knee$ in the distribution at $\sim -1.0$ in the case of the \lalpha{} population and $\sim -0.4$ for the \halpha{} population. Figure \ref{fig:alpha3} reveals that there is a gap in the weighted mean abundance ratios between the two populations. This separation is lower for [Si/Fe] and [Ca/Fe] than for [O/Fe] and [Mg/Fe]. The latter exhibits the largest difference.

It is important to notice that our sample of halo stars detected in APOGEE covers a metallicity range broader than previous work. The \halpha{} population includes objects with $\rm[Fe/H] > -0.4$. Other halo studies had not taken account of stars at larger metallicities because of the difficulty of distinguishing them from disk stars without precise kinematical data. The accurate radial velocities measured in APOGEE allow us to distinguish these halo objects at $\rm[Fe/H] > -0.4$. The halo sample revealed at such high [Fe/H] shows a significant decreasing trend of [X/Fe] with [Fe/H]. This trend was suggested by a handful of objects in NS97 at [Fe/H]$\sim-0.4$, but it is now well-established in this work. 

\begin{figure}
	\includegraphics[scale=0.47, trim=1cm 6cm 2.5cm 7.25cm, clip=true]{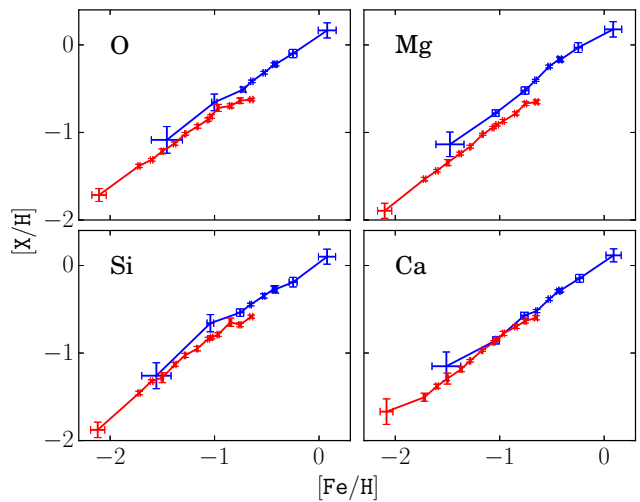}
    \caption{The [X/H] as a function of [Fe/H] for the $\alpha$-elements {O, Mg, Si and Ca in each panel}, obtained from the [X/Fe] weighted means for each stellar population by subtracting the corresponding [Fe/H] mean. Lines represent the abundance ratio means in the "thigh". The populations are color coded as in Figure 4. }
    \label{fig:alpha_h}   
\end{figure}

This broad range in metallicity reveals the $knee$ for both populations. This fact allows us to compare the level of [X/Fe] in stars from each population formed before and after the main contribution of SNIa to the ISM. Consequently, we are able to well-establish whether the \halpha{} population is in fact $\alpha$-enhanced relative to the \lalpha{} population. Since the contribution by SNIa of the $\alpha$-elements Si and Ca is larger than that of Mg, we want to confirm that the enhancement observed in the [X/Fe] vs. [Fe/H] space is also detectable in [X/H] vs. [Fe/H]. From the weighted means we obtain the $\alpha$-to-hydrogen ratios by subtracting the corresponding mean [Fe/H]. Figure \ref{fig:alpha_h} shows the [X/H] mean abundances as a function of [Fe/H] for each population in the "thigh". We see that the \halpha{} population reaches higher [X/H] values than the \lalpha{} population at [Fe/H] $\sim -1$. In addition, the former has its $knee$ shifted to a higher [Fe/H] and includes stars with higher [Fe/H] than the latter (which do not show stars at $[Fe/H] > -0.6$ dex -- see Table 1). This implies that the \halpha{} population is also metal-enriched relative to the \lalpha{} population.

As detected in previous works, we see that the separation between the two populations depends on the element considered. Besides, although all the APOGEE-reliable $\alpha$-elements show a decrease of [X/Fe] with [Fe/H], the slope differs from one element to another, specially at the lowest metallicities. This is however expected, since the yields from the very massive progenitor stars for these very metal-poor objects are different for each element. However, the slope of [Mg/Fe] with [Fe/H] from the $knee$ up to $[Fe/H]_{h}$ in the \lalpha{} is less steep than the slope at the same range in metallicity observed for [O/Fe]. This is not expected at all, considering the current yields of Mg and Fe for SNIa, which predict a very low contribution of [Mg/H] and [O/H], and greater contributions of [Si/H] and [Ca/H]. Consequently, the slope would be similar to that for [O/Fe] and steeper than that observed for [Si/Fe] and [Ca/Fe]. This is not what we observed from our \lalpha{} chemical trends. [O/Fe], [Si/Fe] and [Ca/Fe] exhibit similar slopes, which are steeper than for [Mg/Fe].

\subsection{
Inferences from the Chemical-Evolution Models}


As explained in Section 3.1, the star-formation parameters are inferred by studying the two metallicity ranges: "thigh" and "shin".


\subsubsection{Upper mass limit for the IMF, empirical yields and star formation efficiency}

We derived the $M_{up}$ values for each halo population as described in Section 3.1 and Figure \ref{Mup}. 
We obtain that $M_{up}$ for the \halpha{} population ($26.4 \pm 1.3 M_{\odot}$) is higher than for the \lalpha{} population ($17.9 \pm 2.7 M_{\odot}$). Subsequently, the fraction $f_{SNII}$ derived for the \halpha{} data is higher than that inferred from the \lalpha{}. The resulting values for both parameters are shown in Table \ref{tab:mup_sn2}. The fact that $M_{up}^{\halpha{}} > M_{up}^{\lalpha{}}$ implies that the \halpha{} population formed from an ISM polluted by more massive stars than the \lalpha{} population.

From the derived $M_{up}$ values, and fixing the average $Fe$ yields for massive stars, we obtain the correction factors for $\alpha$-elements that should be applied to the theoretical yields to fit the abundance ratios. These are shown in the upper panel of Figure \ref{fig:1k}. We find that they are fairly well-approximated by unity. Thus, the derived $M_{up}$ ranges are representative of the true $M_{up}$ of each population. 

In general, the correction factors for \lalpha{} are slightly higher than for the \halpha, except for Mg. The difference between the Mg correction factors is the largest, due to the fact that the [Mg/H] difference between the \halpha{} and the \lalpha{} is the highest among the four $\alpha$-elements (see Figure \ref{fig:alpha_h}).

We also derive the star-formation efficiency, $\nu$, for each stellar population from each of the $\alpha$-elements (see Eq. 8).
The bottom panel in Figure \ref{fig:1k} depicts the $\nu$ values. 
The efficiencies are in excellent agreement within the results from each elemental-abundance ratio. Besides, the $\nu$ for the \halpha{} population is higher.


From subsection 3.1 we infer the SFR as a function of time. Figure \ref{fig:sfr} shows the resulting SFR for both populations, as a function of time on the left, and as a function of [Fe/H] on the right.
The equivalence between $t$ and [Fe/H] is given by Eq. (7) and (8). 



The SFR for the \halpha{} population is higher during most of the evolution and decreases more steeply than for the \lalpha{} population, because $\nu_{HMg} > \nu_{LMg}$. The time at which the star formation ends is lower for the \lalpha{} population, meaning a shorter SFH.
  In conclusion, our results imply that the \halpha{} stars formed from a more efficient and longer SFH than the \lalpha{} population. 

\begin{table}
\renewcommand{\thetable}{\arabic{table}}
\centering
\caption{Upper mass limit for the IMF determined from the observed [X/Fe] (see Fig. \ref{Mup}), and the subsequent $f_{SNII}$}\label{tab:mup_sn2}
\begin{tabular}{ccc}
\tablewidth{0pt}
\hline
\hline
Population & $M_{up} (M_\odot)$ & $f_{SNII}$ ($10^{-3}$)\\
\hline
\decimals
 \halpha{}  &  $26.4 \pm 1.3$   & $4.59 \pm 0.1$ \\
\hline
 \lalpha{}  &  $17.9 \pm 2.7$  & $4.00 \pm 0.4$  \\
\hline\hline
\end{tabular}
\end{table}


\subsubsection{Contribution of SNIa and SNII in the "shin"}


It is well known that the steeper downward slope of [X/Fe] vs. [Fe/H] beyond the $knee$ is due to the contribution of SNIa. 
Therefore, we derive their contribution, taking into account the SNIa yields and the empirical yields for SNII, 
the latter $Z$-averaged at the metallicities in the "shin". 

The upper panel of Figure \ref{fig:ka} shows the fraction of SNIa that occurs in $1 M_{\odot}$ of stellar mass. The resulting values, $\sim 10^{-3}M_{\odot}$, are on the order of the $f_{SNIa}$ observed in dwarf galaxies (Maoz et al. 2008). The results obtained from the four $\alpha$-elements show different trends between the populations. However, our results suggest that there is not a significant difference between the fraction of SNIa that contributes to each population. The differences between populations are within the typical errors found in dwarf galaxies.
The low $f_{SNIa}$ value obtained from Mg is due to the flatter slope in the "shin" for the \lalpha{} population.
The resulting $f_{SNIa}$ from O shows a larger value for the \lalpha{} population, because the slope of its "shin" is steeper than for 
the \halpha{} population.



Figure \ref{fig:ka2} exhibits the percentage contribution of SNII and SNIa to the $\alpha$-element enrichment during the "shin". As expected, in this metallicity range the main contribution to O and Mg is due to SNII, whereas Si and Ca have important contributions ($\sim35\%$ and $\sim15\%$, respectively) from SNIa.





\begin{table}
\renewcommand{\thetable}{\arabic{table}}
\centering
\caption{Inferred times at which the highest [Fe/H] occurred for each population, if we consider $t_k= 1$ Gyr.}\label{times}
\begin{tabular}{ccc}
\tablewidth{0pt}
\hline
\hline
 Population & element & $t_h$ (Gyr) \\
\hline
\decimals
 \halpha{} & O  &  2.90 \\
   & Mg  & 2.48  \\
   & Si  & 1.86  \\
   & Ca  & 2.49  \\
   & Fe  & 2.31  \\
\hline
 \lalpha{} & O  & 1.87  \\
   & Mg & 2.32  \\
   & Si & 1.64  \\
   & Ca & 1.95  \\
   & Fe  & 1.92 \\
\hline\hline
\end{tabular}
\end{table}

\section{Discussion}

We have explored and modeled the chemical evolution of the two halo populations seen clearly in APOGEE data, as described in Paper I. 

Halo stars selected by their large radial velocities and a non-disk-like motion in the GRV/cos($b$) vs. $l$ space within the APOGEE DR13 database cover a metallicity range $-2.2 <$ [Fe/H] $< +0.5$. This broad range in metallicity reveals the $knee$ for the two halo populations. The metal-rich side of the \halpha{} population was unexplored in previous works (Nissen \& Schuster papers; Hawkins et al 2015; Paper I). The population with higher $\alpha$-to-iron ratios was truncated at -0.4 or lower metallicities in their samples, and the high-[$\alpha$/Fe] trend with metallicity was described as flat and constant. Our work reveals that this population also shows a steeper decreasing trend at higher metallicities, similar to the decrease observed in the NS10 $low-\alpha$ population, which they ascribed to an increase of iron abundance in the ISM from the contribution of Type Ia supernovae. 

NS10 and later works tried to explain the chemical differences observed between populations in the metallicity range -1.6 $< [Fe/H] <$ -0.4 in terms of the contribution of SNIa for the $low-\alpha$ population due to a lower SFH. The SFH of the $high-\alpha$ population should have been faster in order to reach larger metallicities without the contribution of SNIa. Kobayashi et al. (2014) pointed out that these chemical differences observed between populations could be accounted for by yields from massive stars between 10 and 20 solar masses. They suggested that there could be a difference in the IMF which led to the differences detected in the chemical abundances.

However, it is important to notice that they were trying to explain chemical differences in objects that would have formed from an ISM enriched before and after the pollution by SNIa (the $high-\alpha$ and $low-\alpha$ population, respectively). Since we detect the metallicity value at which the contribution of SNIa became relevant, we are able to compare those objects from each stellar population which had the same kind of progenitors. Therefore, we are able to clarify their hypothesis.

On the one hand, we see that the metallicity range analyzed by Nissen \& Schuster (-1.6 $< [Fe/H] <$ -0.4) comprises objects before the $knee$ for the \halpha{} population and before and after the $knee$ for the \lalpha{} population. This fact implies that the differences observed are, at least partially, due to the contribution of SNIa, as NS10 suggested. On the other hand, the inference of the IMF upper mass limit from objects before the $knee$ for both populations lets us ascertain whether there is, in addition, a difference in the IMF between the populations. We obtain that there actually is a difference in the IMF. Besides, we derive that the upper mass limit for the \lalpha{} population is between 10 and 20 solar masses, as pointed by Kobayashi et al. (2014). Therefore, we conclude that the chemical differences previously detected by NS are due to the combination of a difference in the IMF as well as the contribution of SNIa for the \lalpha{} stars at metallicities lower than -0.4, at which point there was not yet a contribution of SNIa for the \halpha{} population.   

The parameters inferred from these two different chemical trends lead us to two populations with different SFHs:
\begin{enumerate}
\item One population with an IMF weighted to more massive stars, and an SFR more intense and extended in time, and
\item A second population with a top-lighter IMF, and a lower and shorter SFH.
\end{enumerate}

The SFR($t$) from the \halpha{} stars behaves similarly to that of the inner Galactic disk ($r \sim 4$ kpc) during the last 10 Gyr, while the SFR($t$) from the \lalpha{} stars resembles that of the intermediate Galactic disk ($r \sim 8$ kpc) during the last 7 Gyr (Carigi \& Peimbert 2008). Both regions of the Milky-Way disk are explained assuming an inside-out scenario (see Carigi \& Peimbert 2008, 2011). Therefore, we also may explain the two halo populations as resulting from an inside-out scenario for halo formation, where first the \halpha{} stars formed in the inner halo, and immediately after the \lalpha{} stars formed in the outer halo. The IMF for the inner halo needs to be top-heavier to match its $\alpha$-enhancement, and the outer halo requires a dynamically disrupted component to reproduce the retrograde mean orbit observed by NS10.


On the other hand, our results are also consistent with massive satellites reaching and populating more inner regions within the host galaxy during its formation (Tissera et al. 2014; Amorisco 2017). The more massive the satellites are able to continue star formation after they enter the virial radius of the host galaxy. This implies that their SFR would be more extended in time. The less massive satellites would not be able to survive inside the galactic potential, which means that they would be likely to populate the outer regions. They will be disrupted and their star formation would cease. Their SFR before the disruption would be lower because of their lower masses.


Recent results have shown that the top-mass end of the IMF may vary
from galaxy to galaxy and across the galaxies to explain the
dark-matter and baryonic mass-to-light ratio (Capellari et al. 2012;
Lyubenova et al. 2016). Even more, variations are found also across
each galaxy, with clear correlations with the stellar metallicity
(Martin-Navarro et al. 2015). An IMF dominated at early times by high-mass stars would also produce an enhanced
[Mg/Fe] (Mart\'in-Navarro et al. 2015), presenting old stellar populations and being
produced by a strong and short star-formation event (Walcher et al. 2015). This agrees with our results
indicating that HMg stars were formed in a stronger star-formation
event, with a shorter decline time than LHg ones.

\begin{figure}
\includegraphics[scale=0.48, trim= 1.3cm 6cm 0.5cm 7cm, clip=true]{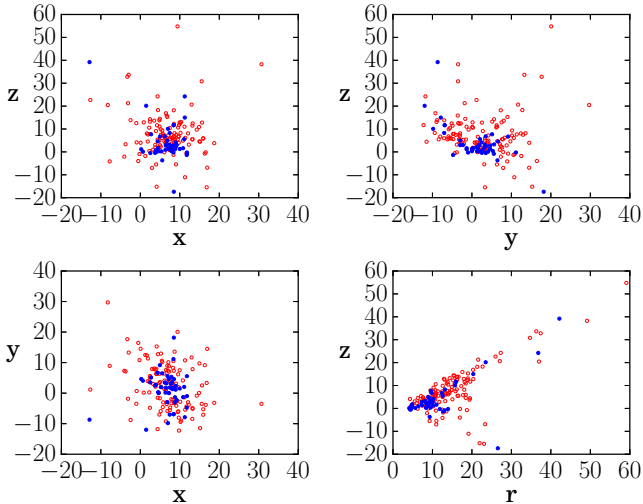}
\caption{Spatial distribution of our sample of stars in the
Cartesian reference system centered in the Galactic center, where $x$ is positive toward the  Sun (considered at a distance of 8.5 kpc), $z$ is positive toward the North Galactic Pole, and $y$ positive toward the direction of Galactic rotation (top panels and bottom left panel). The bottom right panel shows the distance from the Galactic plane ($z$) as a function of the distance from the Galactic center ($r$). The populations are color-coded as in Figure \ref{fig:mgfe}.}
\label{fig:xyz}
\end{figure}

Figure \ref{fig:xyz} shows the space distribution of our sample, using distances from the BPG (Santiago et al. 2016). From the xy, yz and xz planes (in Galactic coordinates), we see that the \halpha{} population is mainly confined at inner regions of the halo. The bottom right panel shows the distance from the Galactic plane, z, as a function of distance from the Galactic center, $r$. The \halpha{} stars are concentrated nearer the Galactic plane ($|z| \leq 5$ kpc) and the \lalpha{} stars reach to larger distances from the center of the Galaxy ($r \geq 15$ kpc) and from the Galactic plane. This is consistent with our previous conclusions.

The abundance dispersion is larger for the \lalpha{} population than for the \halpha{} population. It is also larger than the errors in the measurements. This fact suggests that this \lalpha{} sample is comprised of several populations, i.e., stars formed from environments with different previous enrichments. Moreover, the dispersion may also be caused by the stochastic pollution by massive stars, as the stochastic effects are more relevant when massive stars die in a metal-poor ISM inside small satellites (Carigi \& Hern\'andez 2008).

We assume a simple chemical evolution model which is able to reproduce the chemical trends observed. We do not need to claim for inflows or outflows. However, a kinematical analysis with the precise data provided in the following Gaia data releases will help to reveal the origin of the stars and better clarify the stellar populations comprised in our sample and their chemical trends. A more complex chemical evolution model might be necessary then. It is also necessary to establish with simulations the accretion history of the Galaxy, and better establish whether the chemical trends observed in halo stars could be the result of an inside-out scenario or different kinds of satellites accreted at different times from the host halo of the Galaxy.

\section{Summary and conclusions}

We evaluate chemical trends in the [X/Fe] vs. [Fe/H] space from 175 stars selected within the DR13 APOGEE database, at $4000 < T_{\rm eff} < 4500$ K, for which abundances of the $\alpha$-elements O, Mg, Si and Ca calculated by ASPCAP have the highest accuracy (mean uncertainties $\sim 0.05$ dex). We infer the IMF upper mass limit, fractions of SNII and SNIa relative to the total stellar population, and the star formation efficiency, following a closed-box model of the chemical evolution of the Galaxy under a semi-instantaneous recycling approximation. 

We obtain that: 

\begin{enumerate}
\item Two populations are distinguishable for each $\alpha$-element in the [X/Fe] vs. [Fe/H] space. Their [$\alpha$/Fe] vs. [Fe/H] trends are in agreement with those found by NS10 and NS11 for halo stars.

\item The metallicity range covered extends, at both the low- and high-metallicity limits, beyond that analyzed for the two halo populations previously. This is the first time that these two populations are analyzed over the range $-2.2 < \rm[Fe/H] < +0.5$. 

\item Both populations exhibit a decreasing $\alpha$-to-iron abundance trend associated with Fe enrichment of the ISM by SNIa. This change in the slope (at -0.4 and -1.0 for the \halpha{} and \lalpha{} populations, respectively) is also observed for all the $\alpha$-elements examined, except Ti.

\item Thus, we compare stars before the $knee$ and beyond the $knee$, i.e., objects formed from an ISM with the contribution of only SNII and objects from an ISM with the additional contribution of SNIa. This permits a proper comparison between both populations, in order to clarify whether one population is $\alpha$-enhanced with respect to the other. We corroborate that the population with higher $\alpha$-to-iron values revealed by NS10 is in fact $\alpha$-enhanced with respect to the other. Besides, this population is also metal-enriched respect to the \lalpha{} population.


\item According to our closed-box model, more massive stars contribute to the ISM where the \halpha{} formed with respect to the \lalpha{} population, which implies an IMF weighted to a higher upper mass limit.

\item There is no significant difference between the two populations regarding the contribution of SNIa to enrich the ISM from which the populations formed.

\item The star-formation rate was higher in the \halpha{} population, decreases more steeply with time, and was longer than the SFR(t) inferred for the \lalpha{} population. The latter was lower at early times, more constant, and shorter. 
\end{enumerate}

\section*{Acknowledgements}

E.F.A. acknowledges support from DGAPA-UNAM postdoctoral fellowships. L.C. thanks for the financial supports provided by CONACyT of M\'exico (grant 241732), by PAPIIT of M\'exico (IG100115, IA101215, IA101517) and by MINECO of Spain (AYA2015-65205-P). W.J.S. thanks for the financial supoorts provided by PAPIIT of M\'exico (IN103014). C.R.H. and S.R.M. acknowledge NSF grants AST-1312863 and AST-1616636. C.A.P. is thankful to the Spanish Government for funding for his research through grant AYA2014-56359-P. 
T.C.B. acknowledges partial support from grant PHY 14-30152; Physics Frontier Center/JINA Center for
the Evolution of the Elements (JINA-CEE), awarded by the US National Science Foundation. D.A.G.H.  was  funded  by  the Ramón  y  Cajal  fellowship  number  RYC-2013-14182. D.A.G.H.  and  O.Z.  acknowledge  support  provided  by the Spanish Ministry of Economy and Competitiveness (MINECO)  under  grant  AYA-2014-58082-P. B.T., J.F.-T. and D.G. gratefully acknowledge support from the Chilean BASAL Centro de Excelencia en Astrof\'isica
y Tecnolog\'ias Afines (CATA) grant PFB-06/2007. S.V. gratefully acknowledges the support provided by Fondecyt reg. 1170518.

Funding for the Sloan Digital Sky Survey IV has been provided by
the Alfred P. Sloan Foundation, the U.S. Department of Energy Office of
Science, and the Participating Institutions. SDSS-IV acknowledges
support and resources from the Center for High-Performance Computing at
the University of Utah. The SDSS web site is www.sdss.org.

SDSS-IV is managed by the Astrophysical Research Consortium for the 
Participating Institutions of the SDSS Collaboration including the 
Brazilian Participation Group, the Carnegie Institution for Science, 
Carnegie Mellon University, the Chilean Participation Group, the French Participation Group, Harvard-Smithsonian Center for Astrophysics, 
Instituto de Astrof\'isica de Canarias, The Johns Hopkins University, 
Kavli Institute for the Physics and Mathematics of the Universe (IPMU) / 
University of Tokyo, Lawrence Berkeley National Laboratory, 
Leibniz Institut f\"ur Astrophysik Potsdam (AIP),  
Max-Planck-Institut f\"ur Astronomie (MPIA Heidelberg), 
Max-Planck-Institut f\"ur Astrophysik (MPA Garching), 
Max-Planck-Institut f\"ur Extraterrestrische Physik (MPE), 
National Astronomical Observatories of China, New Mexico State University, 
New York University, University of Notre Dame, 
Observat\'ario Nacional / MCTI, The Ohio State University, 
Pennsylvania State University, Shanghai Astronomical Observatory, 
United Kingdom Participation Group,
Universidad Nacional Aut\'onoma de M\'exico, University of Arizona, 
University of Colorado Boulder, University of Oxford, University of Portsmouth, 
University of Utah, University of Virginia, University of Washington, University of Wisconsin, 
Vanderbilt University, and Yale University.












\listofchanges

\end{document}